\documentclass[preprint,12pt]{aastex}

\def\rfe{R$_{\rm FeII}$}
\def\feiiq{\rm Fe{\sc ii }$\lambda$4570\/}

\def\ltsima{$\; \buildrel < \over \sim \;$}
\def\simlt{\lower.5ex\hbox{\ltsima}}            
\def\gtsima{$\; \buildrel > \over \sim \;$}

\def\simgt{\lower.5ex\hbox{\gtsima}}            

\def\rk{{$R{\rm _K}$}\/}
\def\civ{{\sc{Civ}}$\lambda$1549\/}

\def\cm3{cm$^{-3}$\/}
\def\hb{{\sc{H}}$\beta$\/}

\def\hbbc{{\sc{H}}$\beta_{\rm BC}$\/}

\def\o4363{{\sc{[Oiii]}}$\lambda$4363\/}

\def\feii{{Fe\sc{ii}}$_{\rm opt}$\/}
\def\fe{{\sc{Fe}}\/}

\def\fe76087{{\sc [Fe vii]}$\lambda$6087\/}

\def\kms{km~s$^{-1}$}

\usepackage{graphicx}
\input epsf.sty

\shorttitle{Radio-loud AGN} \shortauthors{ Sulentic et al. }

\begin{document}


\title{Radio Loud AGN in the Context of the Eigenvector 1 Parameter Space
   \altaffilmark{1}}


\author{J. W. Sulentic\altaffilmark{2},   S. Zamfir\altaffilmark{2}, P. Marziani
\altaffilmark{3}, R. Bachev\altaffilmark{2}, M.
Calvani\altaffilmark{3} D. Dultzin-Hacyan\altaffilmark{4}}

\altaffiltext{1}{Based in part on data collected at ESO La Silla.}
\altaffiltext{2}{Department of Physics and Astronomy, University
of Alabama, Tuscaloosa, AL 35487, USA}
\altaffiltext{3}{Osservatorio Astronomico di Padova, Vicolo
dell'Osservatorio 5, I-35122 Padova, Italy}
\altaffiltext{4}{Instituto de Astronom\'\i a, UNAM, Apdo.Postal
70-264, 04510 Mexico D.F., Mexico}


\begin{abstract}
We consider the properties of radio-loud (RL)  AGN in the context of the
Eigenvector 1 (E1) parameter space. RL sources show a restricted E1 parameter
space occupation relative to the radio-quiet (RQ) majority. The Fanaroff-Riley
II ``parent population'' of relatively un-boosted RL sources (median
radio/optical flux ratio $\sim$490) shows the most restricted occupation.  RL
sources have different broad line properties (and inferred black hole masses
and Eddington ratios). FWHM \hb\ for the broad line component in RL sources
are at least twice as large as the RQ majority. The average broad \feiiq\
emission line strength is also about half that for RQ sources. Our sample
suggests that the RL cutoff occurs near \rk $\approx$ 70 or log $P_{\rm 6cm}
\sim 32.0$ ergs s$^{-1}$ Hz$^{-1}$. Sources below this cutoff are RQ although
we cannot rule out the existence of a distinct intermediate population. We show
that the Doppler boosted core-dominated RL sources (median flux ratio
$\sim$1000) lie towards smaller FWHM(\hbbc) and stronger \feii\ in E1 as
expected if the lines arise in an accretion disk. Our subsample of
superluminal sources, with orientation inferred from the synchrotron self
Compton model, reinforce this general E1 trend and allow us to estimate the
role of source orientation in driving E1 domain occupation.
\end{abstract}


\section{Introduction}

We recently presented an expanded spectroscopic database for 215 low-$z$\ AGN
(Marziani et al. 2003a). The moderate s/n and resolution spectra cover the
rest frame region $\approx$ 4300--5300 \AA. Our expanded sample allows us to
address a number of RL issues that have been much discussed recently. We note
that  data with lower resolution or s/n {\em cannot} address these issues in
an Eigenvector 1 (Sulentic et al. 2000ab) context because the uncertainties in
line parameter measures will blur the trends that we find. An important issue
involves claims that a radio-loud (RL) vs radio-quiet (RQ) dichotomy does {\em
not} exist (e.g. White et al. 2000, Cirasuolo et al. 2003) in measures of
radio strength (i.e. radio power  or radio/optical flux ratio). If distinct RQ
and RL populations exist,  quasar samples that use  a radio selection
criterion (e.g. FIRST) will be biased toward  the bright end of the radio
luminosity  function (RLF) for RQ quasars. Therefore a distribution of radio
strength measures may  not appear bimodal but simply with a small RL bump on
the wing of the RQ source distribution.  However, even if the distribution is
not intrinsically bi-modal, and RQ and RL sources show a smooth, partly
overlapping RLF, the RL/RQ dichotomy is appreciable in other properties.

\section{The Radio-loud Sources in Eigenvector 1}

There are two alternate approaches which may provide a much clearer answer to
the issue of RQ-RL bimodality; (1) the Eigenvector 1 (E1) parameter space
(Sulentic et al. 2000a,b), and (2) considerations of radio source morphology in
a classical unification scenario. The E1 diagram has the advantage that \rfe\
and FWHM \hb\ are not strongly dependent on luminosity up to at least M$_B$=
-26 (Marziani et al. 2003a; correlation coefficients are $\la$0.1). Therefore
we can reasonably compare RQ and RL subsamples with different luminosity
distributions in this parameter plane.  Figure 1 shows the distribution in the
optical E1 plane for all sources in our sample with log \rk$>$1.0 (where \rk\
is the radio/optical flux ratio as defined in Kellermann et al. 1989). The bulk
of our RL sources are concentrated in the region we call Population B. Our RL
sample is ``unedited" and therefore shows a few sources with \rfe$>$0.4. These
sources involve: (1) our noisiest spectra, (2) sources with extremely broad and
very low EW H$\beta$\ broad component [W(\hbbc) $\sim$20 \AA] with a high upper
limit for \rfe, (3) a few luminous radio core-dominated sources (circled) that
appear to be genuinely \feii\ strong and RL. The bulk of RL sources are
restricted to the parameter ranges 3000$\leq$FWHM(\hbbc) $\leq$10000 \kms\ and
0$\leq$\rfe$\leq$0.4.

Can the RL occupation in E1 be ascribed to sample selection effects?
Systematic differences in FWHM \hb\ between low redshift RQ and RL AGN are
confirmed in samples where $z$\ and apparent magnitude distributions are
matched (Marziani et al. 2003b). Perhaps a large population of RL sources that
will fill in E1 remains to be discovered, at least to the point of removing any
difference in domain space occupation. While our sample is incomplete, it is
certainly sampling a significant fraction of both RL and RQ sources brighter
than V=16.5 (see Figure 2 in Marziani et al. 2003a).
It is unlikely that we have missed a major RL population
with significantly different properties. At the same time, growing samples of
narrow line Seyfert 1 sources (e.g. Grupe et al. 1999) (we would call them
extreme RQ pop. A) contain almost no RL sources (see later discussion). The
different RL domain space occuption in E1 argues in favor of a RL-RQ
bimodality.

Figure 1 gives us  more information because we have also examined the radio
morphologies of all sources in our sample with log \rk$\la$1.0. The
characteristic radio morphology of a classical RL source involves double-lobed
structure centered on a host galaxy (at low redshift) showing broad nuclear
emission lines that mark it as hosting a type 1 AGN. Whether the central host
can be seen or not, varying degrees of core radio emission are seen between the
double lobes with jet structure often connecting core and lobes. Double lobed
morphology in the context of type 1 AGN usually means Fanaroff-Riley (1974)
class II sources (FRII). We identify 39 FRII sources in our RL sample (7 are
technically HYMOR following Gopal-Krishna \& Wiita 2000). This is somewhat
more than half of our RL sample (62 with log \rk$>$2.0). The remaining RL
sources in our sample show core or core-jet radio structure. In a classical
morphology unification scenario they are interpreted as double-lobed sources
aligned to within $\approx$25$^\circ$ of  our line of sight. In order to keep
our definitions simple our core-dominated (CD) sources show no evidence for
double-lobed structure in any published radio map (this will depend on the
spatial frequency sensitivity and dynamic range of available maps).

We interpret FRII sources as the parent population of RL AGN. They are shown
as open circles in Figure 1 and the vast majority show a strong concentration
in the population B domain of E1. These are unambiguously RL sources and all
show log \rk $\geq$1.8--we find three additional FRII sources with 1.8$\leq$log
\rk $\leq$2.0. Until weaker FRII sources are found we adopt log \rk $\sim$1.8
as the lower limit for classical RL sources (the three open circles with log
\rk $>$1.0 on the right side of figure 1 will likely move into the pop. B
region with better data). We make this statement because the CD RL analogues
to the FRII sources are interpreted as preferentially aligned sources. Their
lobe/jet axis should be aligned within a few degrees to our line of sight. One
observes rapid variability at many wavelengths, apparent superluminal motion
in the radio structure and Doppler  boosting of the continuum emission (even
swamping the emission lines in Blazars). The latter effect is most relevant
here, because in the simplest unification the CD RL sources should all be
Doppler boosted. while  the FRII sources will be unboosted or less boosted
than CD sources. The same source viewed pole-on will show a larger \rk\ value
(dependent on the boosting factor,
 Ghisellini et al. 1993) than when viewed lobe-on. CD sources with log \rk\ $>$2.0 are
marked as filled squares in Figure 1. Median log \rk\ values for our FRII and
CD subsamples are 2.7 and 3.0 respectively with large scatter. The restricted
domain space occupation for RL sources, compared to RQ sources, in E1 is
perhaps the strongest argument for interpreting RL sources as being
fundamentally different from the RQ majority of type 1 AGN. At the very least
it implies that they are, in some sense, an extremum in a continuous sequence
of Type 1 AGN broad line region (BLR) properties.

Figure 1 provides even  more information because we can now compare FRII and
CD domain occupation with predictions of the favored models for low ionization
broad (including Balmer) line emission. Our E1 motivated working hypothesis
has been that: (1) highly accreting (as supported by E1 parameters involving a
\civ\ blueshift and a soft X-ray excess), low black hole (BH) mass population A
sources (Sulentic et al. 2000b, Marziani et al. 2001, 2003b) show Balmer and
\feii\ emission from an accretion disk, while,  (2) the larger BH masses and
lower accretion rates estimated for RL + pop B RQ sources (Marziani et al.
2001, 2003b) coupled with the absence of a \civ\ blueshift and soft X-ray
excess implied different disk/wind properties, and probably a single emitting
region (Marziani et al. 2003c). Claims of no RL-RQ mass difference (e.g.
Oshlack et al. 2002) may reflect smaller measured FWHM values resulting from
nonsubtraction of a narrow component from the Balmer lines -- $M_{\rm
BH}$$\propto$[FWHM(\hbbc)]$^2$.

We have suggested that the stronger Balmer line and weaker \feii\ emission from
pop. B RL and RQ sources might be ascribed to thermal gas entrained along the
jet axis (i.e., a bicone). In a bicone dominated emission line scenario we
would expect to see the broadest Balmer profiles in sources oriented near
pole-on. A model viewing line emission from an accretion disk would make the
opposite prediction-- the Balmer lines would be narrowest for near face-on
oriented CD sources (see also Wills \& Browne 1986). We would also expect
stronger \feii\ emission from a line emitting disk when it is viewed face-on
(Marziani et al. 2001). We would therefore expect RL CD sources to be
displaced towards lower FWHM(\hbbc) and larger \rfe\ in a disk scenario. It is
clear from Figure 1 that a disk model is favored since most CD sources are
found at the lower edge of the FRII distribution and displaced towards
slightly higher average \rfe. E1 has enabled us to isolated the parent FRII RL
population and to estimate the role of source orientation in driving RL source
occupation.

\subsection{Superluminal RL Sources in E1}

We find a domain space separation between FRII and CD sources in E1. Following
the  same line of reasoning we can examine the superluminal sources  in our
sample. Jet orientation can be inferred from radio observations that allow
estimation of the synchrotron self Compton flux relative to the observed X-ray
flux (Ghisellini et al. 1993). Using the same reasoning, Rokaki et al (2003)
compared FWHM(H$\alpha$) measures with predictions of jet orientation
($\theta$) and estimated Doppler boosting factor ($\delta$) for a sample of
superluminal sources. They find a weak correlation consistent with the notion
that the Balmer lines arise from an accretion disk whose plane is roughly
perpendicular to the jet axis.

Figure 2 shows our $\theta$\ vs FWHM(\hbbc)  correlation diagram
for 9 superluminal sources in common with Rokaki et al. plus two
additional sources from Marziani et al. (2003a). Our optical
\hbbc\ measures are generally higher resolution and s/n than the
IR H$\alpha$\ measures yielding a lower scatter. We confirm their
conclusion that the correlation is consistent with a disk origin
for the bulk of Balmer broad line emission and that the
correlation is opposite of that expected in a bicone scenario.
The diagonal arrow in figure 1 indicates the average expected
change in E1  position due to change in orientation from FRII to
CD sources. Many of the most extreme boosted sources will be
lineless blazars and hence absent from E1. It is interesting that
BL Lac in low phase showed FWHM(\hbbc) $\approx$ 4000 \kms, close
to the locus of CD points in Fig. 1. Note that, in an effort at
simplification,  we have ignored the radio spectral indices or
core/lobe ratio in this paper. We note that the most extreme
source identified as FRII in Figure 2 (smallest FWHM and log
theta) is strongly asymmetric and core-dominated.

\section{Distribution of RQ and RL Sources in R$_K$: Radio Intermediates?}

Figure 3 plots \rk\ versus the two optical E1 parameters. We again identify the
FRII and CD RL populations as in Figure 1. We extracted 6cm radio fluxes and
optical B magnitudes from Veron-Cetty et al. (2001).  Similar distributions
are found if ones uses NVSS 20cm fluxes or if one uses  radio power instead of
\rk\ for the abscissa. The RL source preference for FWHM(\hbbc)$\geq$4000 \kms\
and \rfe$\leq$0.5 is clearly seen in the plots. We also see a source deficit
between log \rk\ = 1.0 and 2.0. We do not suggest that this is the distribution
that one would expect for any  complete sample. It will never look like this
for reasons  discussed earlier.
RL sources are clearly overrepresented in our sample for complex reasons. The
focus of interest here involves the most radio ambiguous sources between log
R$_K$=1.0 and 2.0. Three of them turned out to be FRII sources that define our
lower boundary for the classical RL sources. In our interpretation, CD sources
with R$_K$ less than the weakest observed FRII sources have no meaning in a RL
classical unification. Since all but three intermediate sources (n=10-marked
as filled triangles in Figure 1) show core structure we interpret  their radio
emission  as  unrelated to RL core/lobe emission. E1 further supports this
interpretation because sources with log R$_K$$<$1.0 show no domain preference,
they are found just as frequently in RQ pop A domain as in the pop B domain
preferred by RL sources.

We have found a significant number of radio ``intermediate" sources ($1 \la
\log$ \rk\ $\la 2$) and seek to account for them.  At least five
interpretations can be proposed.  (1) The overlap between the faint end of the
RLF for RL sources and the bright end of the RLF for RQ sources. This is
supported by the data since three sources show classical FRII structure--they
are the weakest RL sources in our sample at \rk\ $\approx 70$. This is the
simplest explanation and is further supported by  a study  of the 6-20cm
spectral indices (Ivezic et al. 2002) where it is concluded that almost all low
redshift radio-loud AGN are likely to show log \rk\ $\ga $2.0. Assuming a mean
value of log \rk\ $\approx$ -0.05 for RQ sources and a reasonably symmetric
distribution about the mean would place the high log R$_K$\ end of the
distribution near 0.0. Explanation 2 can then be helpful here. (2) Mixed
starburst/AGN sources with radio emission amplified by processes unrelated  to
the RL jet phenomenon. A good place to look would be among  ULIRG sources. We
find many \rk\ values between $\pm$1.0 with the most extreme sources (e.g. Arp
220 and NGC6240) lying between 1.0-2.0. We can identify 4 objects among our 13
intermediate sources that show an IR excess and are interpreted as mixed
AGN/starbursts (Mkn231, IRAS 0759+64, NGC7674 and Mkn 896) consistent with
enhanced radio emission from activity related to star formation. If a
plausible correction ($A_{\rm B}$$\approx$3.0) for internal extinction is
applied e.g. to Mkn 231 then the observed log \rk\ $\sim$1.8 will decrease to
$\sim$0.6. Most of the remaining 6 intermediate sources are too distant for an
IRAS detection below ULIRG level. (3) A  population of RQ sources that show
weak radio jet structure (Falcke et al. 1996, 1996, de Diego et al. 1998,
Gopal-Krishna et al. 2001, Blundell \& Rawlings 2001). This represents another
method to enhance the radio emission from RQ sources. Would it be related or
unrelated to classical RL activity? The domain occupation  in E1  and the lack
of double lobed morphology argue ``unrelated". (4) Population B spiral
galaxies that are similar to elliptical hosted RL sources in BH mass,
accretion rate, E1 parameters and, consequently, BLR structure although with
their radio emission  beamed but somehow muffled. This interpretation is
clearly related, or even identical, to 3; see refs. for 3). Reliable host
galaxy morphology only exist for a handful of our sources but at least two
known spirals in our sample show true RL (log \rk\ $\ga$ 2) activity 3C120
(complex core/jet emission-also called an FRI source) and III Zw 2. (5) Proto-
RL sources which have not yet produced jet/lobe structure and are heavily self
absorbed (O'Dea 1998).

The remaining 6 intermediate sources are more difficult to explain. If we
interpret these CD sources as classical RL AGN then they should be oriented
with the jet axis near our line-of-sight and consequently beamed (Doppler
boosted).  Such low \rk\ values for boosted sources would imply an un-beamed
parent population that is unambiguously RQ. One does not observe classical
double-lobed sources in RQ samples (Kellermann et al. 1994, Kukula et al.
1998). The intermediate sources show Population B properties with mean
FWHM(\hbbc)$\approx$ 8800 \kms, and \rfe$\approx$0.3. These sources deserve
further study in order to see if they conform to any of the possible
explanations proposed above.

Have we so far missed many radio intermediate sources in our growing sample?
The simplest reason would be if a class of low redshift intermediate radio
emitters was shown to be optically fainter on average than low redshift
populations of RQ and RL sources. One must be careful here because any optical
dimming will tend to enhance rather than diminish \rk. That would point toward
a RQ parent population for the dimmed sources by the same argument that radio
weak/intermediate  CD sources viewed as beamed sources, imply a weaker (i.e.
RQ) parent population.

We note that 3 NLSy1 galaxies (not part of our sample) with log \rk\ $\approx$
1.4, 1.5 and 1.8 (Remillard et al. 1986; Siebert et al. 1999; Grupe et al.
2000) have been proposed as CD RL sources. They would fall in the (population
A) intermediate region of Figure 1 where RL sources are rare. They do not fit
in a classical RL unification scheme  but may be candidate beamed RQ sources.
PKS2004-447 (Oshlack et al. 2001) shows narrow Balmer lines and no \feii\
emission and is probably a type 2 AGN. As far as we are aware Zhou et al.
(2003) have found the first genuinely RL NLSy1 source (log \rk $>$3.0)--it
would fit with the three \feii\ strong CD sources circled in Figure 1.

\section{Conclusion}

We find that: (1) E1 parameters support some kind of dichotomy between RL and
RQ AGN, (2) sources with different radio morphology occupy different regions of
the E1 parameter space supporting the unification assumption that true RL CD
sources can be interpreted as almost face-on FRII's and (3) E1 occupation, and
comparison of superluminal source orientation estimates with E1 parameters,
suggest that a significant part of the Balmer and \feii\ emission arises in an
accretion disk.  The latter result suggests that the RL-RQ dichotomy may be
more related to evolution than to fundamentally different BLR structure.


\clearpage

\figcaption{RL sources in the optical plane of the E1 parameter space (open
circles - lobed sources, filled squares - pure core sources). Intermediate RQ
(10$\la$ \rk\ $ \la 100$) sources are plotted as triangles. Sources below the
horizontal dashed line are called Pop. A while those to the right of the
vertical dashed line are called outliers. The thick arrow indicates
displacement between the median \rfe\ and FWHM(\hbbc) values from lobed and
pure core RL sources which provides an estimate of the role of source
orientation in E1. The circle identifies CD sources with \rfe $\ga$ 0.5.
Errors in FWHM and \rfe\ are the average of $2\sigma$\ errors for sources in
our sample.  \label{fig01}}

\figcaption{Balmer line FWHM vs. inferred line-of-sight orientation angle
($\theta$)  for 9 superluminal sources common between our sample and Rokaki et
al. (2003) + 2 additional superluminals (NGC1275 and 3C345) from the sample of
Marziani et al. (2003a). Stars refer to the H$\alpha$\ measures from Rokaki et
al. and open circles/filled boxes our \hbbc\ measures (for lobed and pure core
sources respectively). FWHM(\hbbc) errors as in Figure 1. The most deviant
point (B2 1721+34) shows a peculiar \hbbc\ profile dominated by a VBLR
component - it is not certain to have a classical BLR component (see Sulentic
et al. 2000c). \label{fig02}}

\figcaption{The radio-optical flux ratio \rk\ as a function of the optical E1
parameters  \rfe\ (bottom) and FWHM(\hbbc) (top). CD sources (i.e. no lobes)
are indicated with filled symbols (boxes for RL and triangles for RQ). Open
circles are used for sources showing double-lobe structure. A clear bimodality
is seen with RL sources concentrated above log \rk $\approx$ 2.0, below log
\rfe $\approx$ 0.4 and above FWHM(\hbbc) $\approx$ 4000 \kms. The horizontal
dashed line (upper panel) marks the nominal population A-B boundary in FWHM
H(\hbbc). Vertical lines mark the zone of intermediate sources. The error bar
for \rk\ has been derived from a comparison between NVSS and 6cm fluxes from
V\'eron-Cetty \& V\'eron (2001). Errors in FWHM and \rfe\ as in Figure 1.
\label{fig03}}

\newpage
\plotone{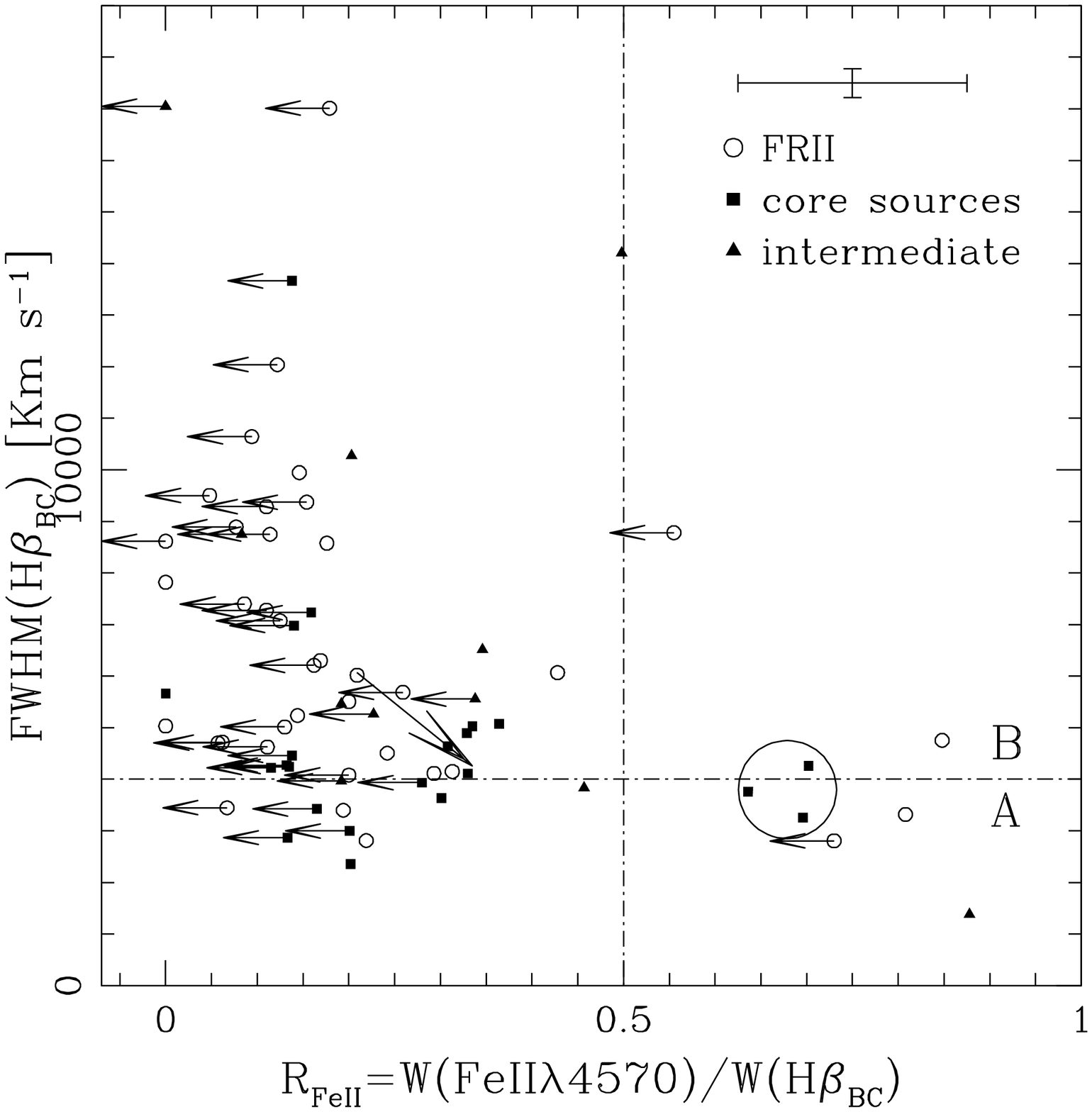}
\newpage
\plotone{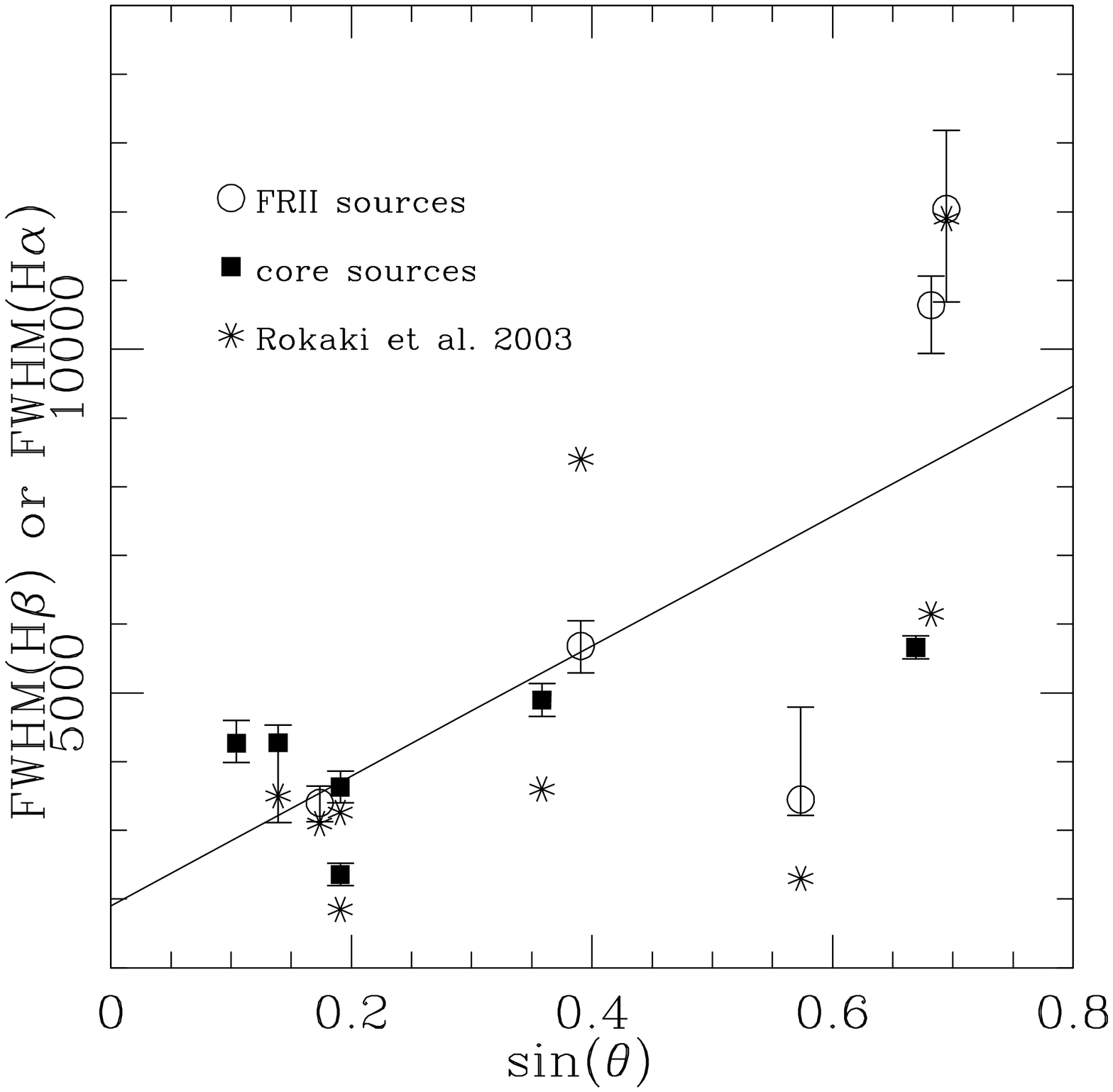}
\newpage
\plotone{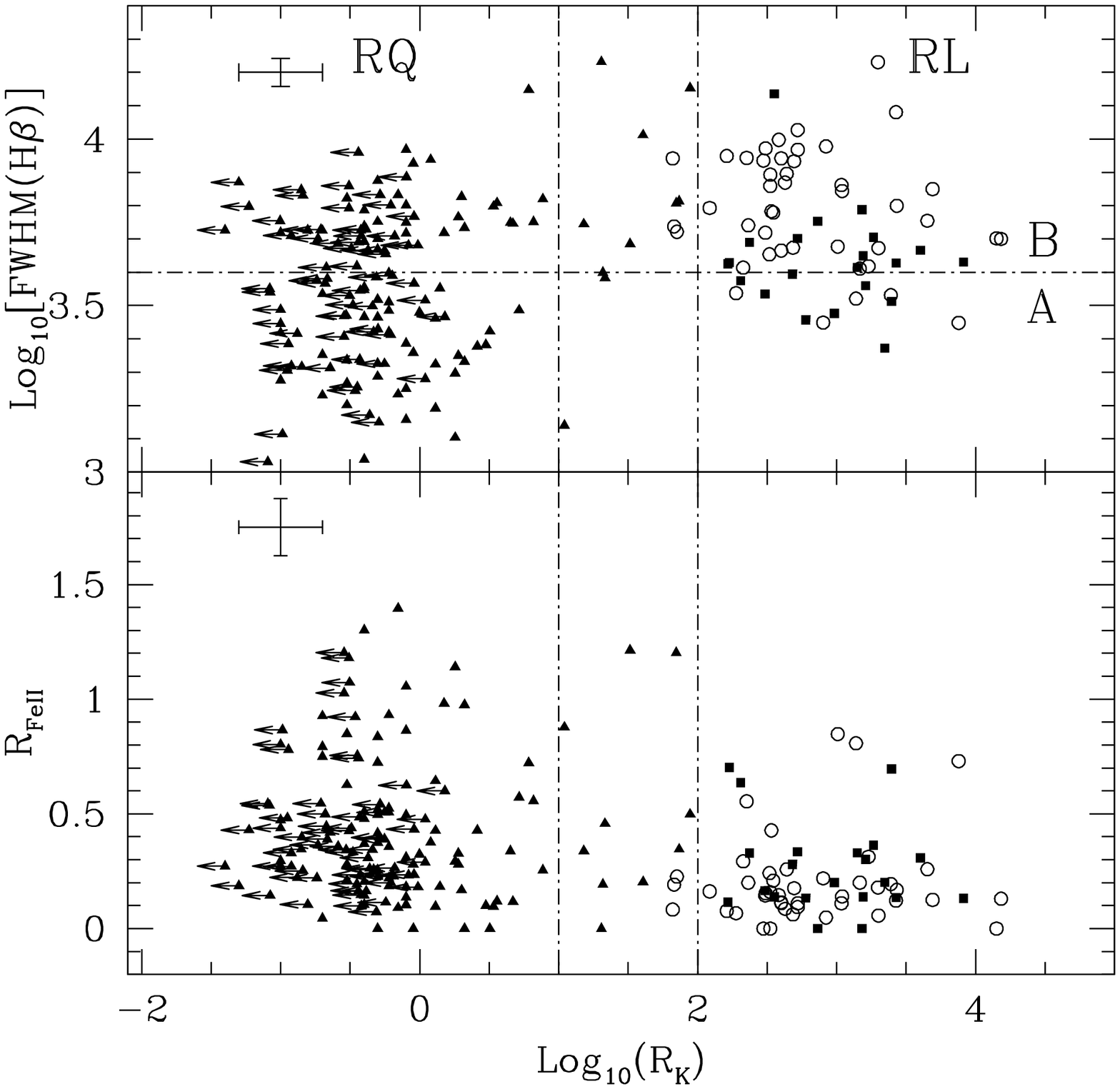}

\end{document}